\documentclass[aps,prb,showpacs,twocolumn]{revtex4}
\usepackage{amssymb}
\usepackage{amsmath}
\usepackage{graphicx}
\usepackage{dcolumn}
\usepackage{bm}
\usepackage{units}
\usepackage{booktabs}
\usepackage{multirow}

\hfuzz=\maxdimen
\begin{document}

\title{Asymmetric angular dependence of spin-transfer torques in CoFe/Mg-B-O/CoFe magnetic tunnel junctions}

\author{Ling Tang}\email{lingtang@zjut.edu.cn}
\author{Zhi-Jun Xu}\email{xzj@zjut.edu.cn}
\author{Xian-Jun Zuo}
\author{Ze-Jin Yang}\email{zejinyang@zjut.edu.cn}
\affiliation{Department of Applied Physics, College of Science, Zhejiang University of
Technology, Hangzhou 310023, China}

\author{Qing-He Gao}
\affiliation{College of Science, Northeastern University, Shenyang 110004, China
\\Information Engineering College, Liaoning University of Traditional Chinese Medicine, Shenyang 110847, China}

\author{Rong-Feng Linghu}\email{linghu@gznu.edu.cn}
\affiliation{School of Physics and Electronics Sciences, Guizhou Education University, Guiyang 550018, China}

\author{Yun-Dong Guo}\email{g308yd@126.com}
\affiliation{College of Engineering and Technology, Neijiang Normal University, Neijiang 641112, China}

\date{\today}

\begin{abstract}
Using a first-principles noncollinear wave-function-matching method, we studied the spin-transfer torques (STTs) in CoFe/Mg-B-O/CoFe(001) magnetic tunnel junctions (MTJs), where three different types of B-doped MgO in the spacer are considered, including B atoms replacing Mg atoms (Mg$_{3}$BO$_{4}$), B atoms replacing O atoms (Mg$_{4}$BO$_{3}$), and B atoms occupying interstitial positions (Mg$_{4}$BO$_{4}$) in MgO. A strong asymmetric angular dependence of STT can be obtained both in ballistic CoFe/Mg$_{3}$BO$_{4}$ and CoFe/Mg$_{4}$BO$_{4}$ based MTJs, whereas a nearly symmetric STT curve is observed in the junctions based on CoFe/Mg$_{4}$BO$_{3}$. Furthermore, the asymmetry of the angular dependence of STT can be suppressed significantly by the disorder of B distribution. Such skewness of STTs in the CoFe/Mg-B-O/CoFe MTJs could be attributed to the interfacial resonance states induced by the B diffusion into MgO spacer. The present investigation demonstrates the feasibility of effectively enhancing microwave output power in MgO based spin torque oscillator (STO) by doping the B atoms into MgO spacer. \end{abstract}

\maketitle

\section{Introduction}

The current-induced spin-transfer torque (STT) can be used to drive a steady magnetization precession,~\cite{Slonczewski-mmm96, Berger-prb96, Kiselev-nat03} which has attracted much attention for applications such as nanometer-sized spin torque oscillator (STO),~\cite{STNO1, STNO2} where a microwave signal with tunable frequency and narrow linewidth can be generated by a direct current (DC). Recently Ikeda \emph{et al.} ~\cite{Ikeda-nm10} found that the CoFeB/MgO/CoFeB magnetic tunnel junctions (MTJs) with interfacial perpendicular magnetic anisotropy (PMA) is a excellent candidate for STO devices. Inspired by this development, the STT induced microwave oscillator based on CoFeB/MgO MTJs, in which the B atoms are introduced during the fabrication of junctions, have been investigated intensively by several experimental researchers. ~\cite{Dijken-ape12, Kanai-apl12, Zeng-sp13, Lebrun-pra14}

However, generating a high emission power in STO devices without external magnetic field still remains a great challenge so far. Rippard \emph{et al.} ~\cite{Rippard-prb10} pointed out that the output power of STO is closely related to the angular dependence of STT, namely the larger asymmetry parameter $\Lambda$,~\cite{Slonczewski-mmm02, jia-prb11} which is proportional to the skewness of STT, the higher output power of STO devices will be achieved. Based on such conclusion of STT induced oscillation, Jia \emph{et al.}~\cite{jia-prl11} predicted a large emission power in Fe/MgO/Fe MTJ with only 3 layers of MgO as spacer, according to the calculated results of strong asymmetric angular dependence of STTs. However, the situation in CoFeB/MgO MTJs, which is widely employed as STO devices, is more complicated than that in Fe/MgO MTJs due to the introduction of B atoms. Therefore, the question about what kind of influence on the angular dependence of STT by introducing B atoms into CoFe/MgO and whether the introduction of B is beneficial to produce high power of microwave emission are still unclear yet.

In addition, some experimental results indicate that the amorphous CoFeB electrodes near the interface in CoFeB/MgO MTJs crystallized into ordered body center cubic (bcc) CoFe crystal after annealing process. ~\cite{Djayaprawira-apl05, Yuasa-apl05, Yuasa-jpd07, Rumaiz-apl10} Meanwhile, the B atoms are pushed out of the CoFeB electrodes and then diffused into the MgO layers forming Mg-B-O spacer.~\cite{Cha-apl09, Lu-jap10, Kurt-apl10, Rumaiz-apl11, Greer-apl12} However, the exact distribution and position of B atoms in Mg-B-O spacer are still unclear up to now. Thus several possible structures of B-doped Mg-B-O spacer have been proposed and investigated. Stewart~\cite{Stewart-nanolett10} suggested that the B atoms diffused into MgO layer forming kotoite Mg$_{3}$B$_{2}$O$_{6}$. Bai \emph{et al.}~\cite{Bai-prb13} studied the transport properties of CoFe/kotoite/CoFe MTJ and concluded that the symmetry reduction is responsible for the lower tunneling magnetoresistance (TMR). In addition, the electronic structures of bulk B-doped MgO with different configurations have been studied~\cite{Liu-jpc10, Chandra-prb14} but the properties of STTs in corresponding CoFe/Mg-B-O based MTJs are still unknown yet. So the purpose of this paper is to investigate which configuration of CoFe/Mg-B-O is most efficient for high microwave output power from the viewpoint of the STT. 

\begin{figure}
  \includegraphics[width=8.6cm]{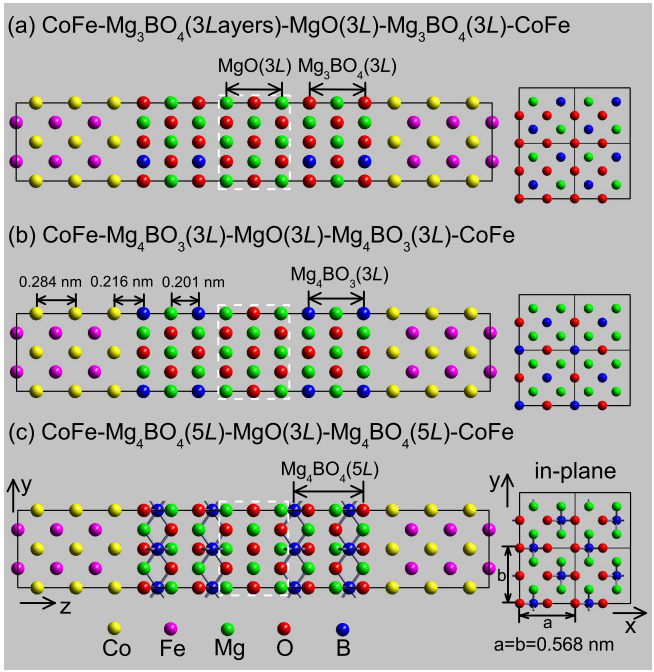}\\
  \caption{(color online). The atomic structures of supercells for three prototypes of CoFe/Mg-B-O/CoFe(001) MTJs. Here we consider three types of dopant in spacer: (a) B substituting for Mg (Mg$_{3}$BO$_{4}$), (b) B substituting for O (Mg$_{4}$BO$_{3}$), and (c) B occupying the interstitial position (Mg$_{4}$BO$_{4}$) in MgO, respectively. The views of Mg-B-O spacer along the transport direction ($z$ axis) are also displayed on the right of figure. The 3 layers (3\emph{L}) undoped MgO in the middle of spacer is introduced to avoid the possible coupling between two ferromagnetic electrodes which might be intermediated by the magnetic B atoms. }\label{structure}
\end{figure}

In this work, we calculate the STT of CoFe/Mg-B-O/CoFe(001) MTJs by noncollinear scattering wave-function matching method in combination with first-principles calculations.~\cite{shuai-prb08, shuai-prb10, tang-ijmp13} Here the B atoms are assumed to diffuse into the MgO layers forming Mg-B-O spacer and three possible types of B-doped (25\%) MgO are considered, i.e., (1) B replacing Mg (Mg$_{3}$BO$_{4}$), (2) B replacing O (Mg$_{4}$BO$_{3}$), and (3) B occupying the interstitial position (Mg$_{4}$BO$_{4}$) in MgO, respectively. Our calculated results show that a significantly asymmetric angular dependence of STT could be obtained in ballistic CoFe/Mg$_{3}$BO$_{4}$ and CoFe/Mg$_{4}$BO$_{4}$ based MTJs, both of which are able to result in the large microwave output power.

\section{Computational Method}

The stuttgart tight-binding linearized muffin-tin-orbital atomic sphere approximation (TB-LMTO-ASA) program~\cite{andersen85} is used to obtain the effective single-electron potential of CoFe/Mg-B-O/CoFe(001) MTJs. A supercell including 11 monolayers (MLs) of ordered bcc CoFe and a spacer partially filled with B-doped MgO is used for such self-consistent calculation. To investigate the influence of B diffusion, three types of B-doped MgO are considered, where B atoms substitute for Mg, O atoms and occupy the interstitial position of MgO. The corresponding formulas of three B-doped MgO can be written as Mg$_{3}$BO$_{4}$, Mg$_{4}$BO$_{3}$ and Mg$_{4}$BO$_{4}$, respectively, where the dopant concentration is 25\%.

\begin{figure}
  \includegraphics[width=8.6cm]{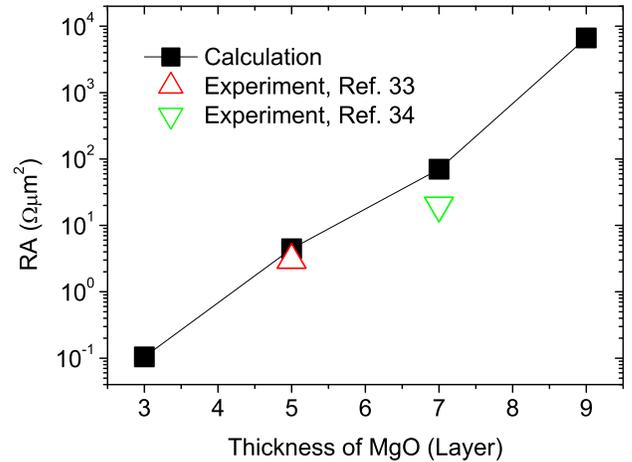}\\
  \caption{(color online). The MgO thickness dependence of resistance-area (RA) in CoFe/MgO/CoFe(001) MTJs. The black squares are our calculated results of ballistic junctions. The red and green triangles are experimental values from Ref. 33 and 34, respectively. }\label{RA}
\end{figure}

\begin{figure}
  \includegraphics[width=8.6cm]{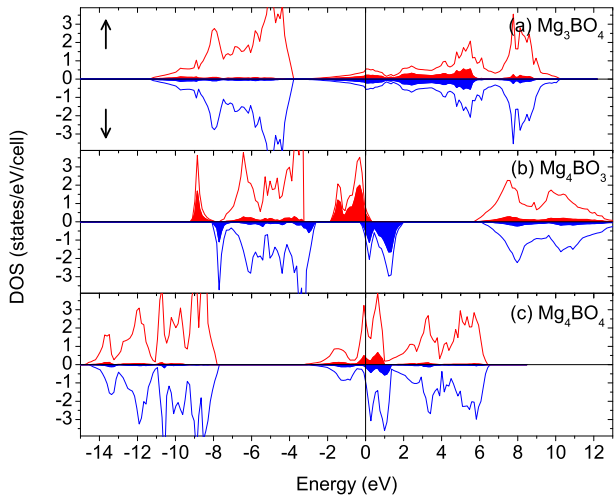}\\
  \caption{(color online). The total DOS (solid) for bulk (a) Mg$_{3}$BO$_{4}$, (b) Mg$_{4}$BO$_{3}$, and (c) Mg$_{4}$BO$_{4}$, respectively. It shows that all three bulk B-doped MgO crystals exhibit metallic state. The projected DOS (shaded) on impurity B atoms are also plotted in (a)-(c), which indicates that at Fermi level the coupling between B and the host MgO is much weaker for Mg$_{4}$BO$_{3}$ than that for the other two types of B-doped MgO. }\label{dos}
\end{figure}

Fig.~\ref{structure} shows the atomic structures of supercells for three prototypes of CoFe/Mg-B-O/CoFe(001) MTJs considered in our calculations, where the B-doped MgO layers locate at the interfacial region near the electrodes and the undoped MgO layers are in the middle of spacer. Here the introduction of middle undoped MgO layers is in order to decouple the possible interaction between pinned layer (the right CoFe lead) and free layer (the left CoFe lead) in CoFe/Mg-B-O/CoFe MTJs. Such interaction might be intermediated by the magnetic B atoms in B-doped MgO [see shaded part in Fig.~\ref{dos}(b) for details]. Moreover, the CoFe lead with ordered bcc phase and Co-terminal interface are used in our calculations, where the Co-O distance at the interface is 2.16  \AA.~\cite{Burton-apl06} The in-plane lattice constant of supercell is $a=b=5.68$ \AA, whereas the distance between adjacent Co and Fe layer is equal to that of bulk CoFe (1.42 \AA). The crystal lattice of B-doped MgO as well as undoped MgO can be matched with CoFe lead in (001) direction by rotating 45 degree around (001) axis and compressing lattice slightly.

With the rigid potential approximation, the STT can be calculated by the noncollinear scattering wave-function in TB-LMTO-ASA basis, where the effective potentials for each atomic spheres are rotated by different angles relative to global quantization axis in spin space.~\cite{shuai-prb08} In addition, those angles are according to the direction of magnetization on each atomic spheres. In the present CoFe/Mg-B-O/CoFe MTJs, the directions of magnetization for the right CoFe lead and magnetic B atoms at the right side of spacer are all align with $y$ axis. For the left CoFe lead and magnetic B atoms at the left side of spacer, the magnetizations are rotated the same relative angle to $y$ axis within $x$-$y$ plane (in-plane). So the transport direction is along the $z$ axis (out-of-plane). In addition, a $240\times240$ $k$-point mesh is used in full two-dimensional (2D) Brillouin zone (BZ) to evaluate the in-plane STTs (T$_{x}$ and T$_{y}$). The out-of-plane torque (T$_{z}$) is excluded in this work due to it needs more $k$ points to ensure convergence, which is beyond our computational capability.

In order to compare the present transport calculation with experiment, the resistance-area (RA) products in ballistic undoped CoFe/MgO/CoFe(001) MTJs are evaluated in this work. Fig.~\ref{RA} demonstrates the calculated RA dependence on the thickness of MgO spacer. The measured RA values from experiments~\cite{Arakawa-apl11, Liebing-jap12} (red and green triangles) are also plotted in Fig.~\ref{RA}. The present calculations show that the RA product decreases exponentially with reducing the thickness of MgO spacer. Moreover, for MTJ with 5 MgO layers ($\sim$ 1 nm thick) as spacer the calculated RA is 4.44 $\Omega\mu\text{m}^{2}$ that is close to the experimental value~\cite{Arakawa-apl11} of 3.01 $\Omega\mu\text{m}^{2}$.

\begin{figure}
  \includegraphics[width=8.6cm]{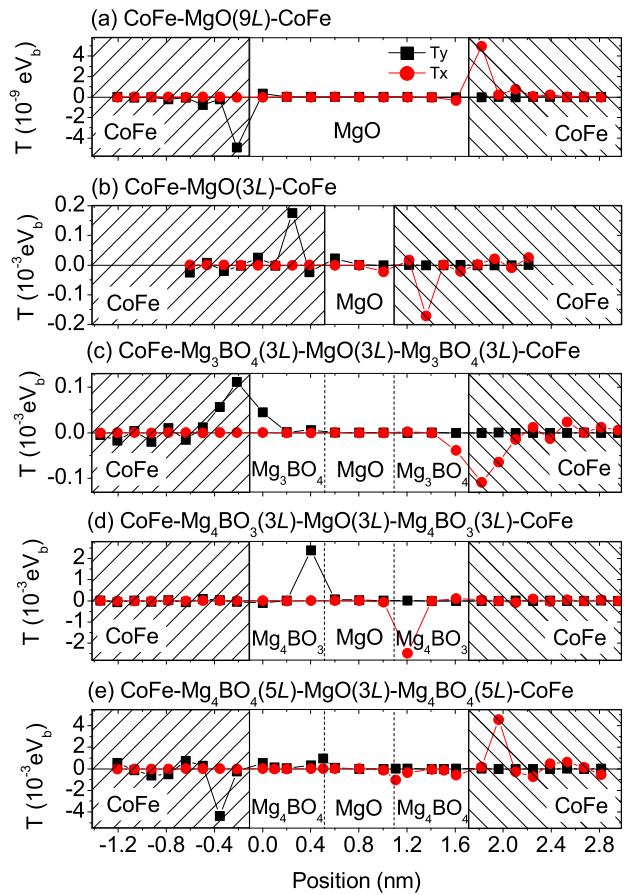}\\
  \caption{(color online). The layer resolved in-plane STTs (T$_{x}$ and T$_{y}$) of MTJs at unit bias with and without B diffusion for the relative angle of 90 degree between two leads. It shows that the torques mainly distribute around the CoFe/Mg-B-O interface except for the case of CoFe/Mg$_{4}$BO$_{3}$ based MTJ. As shown in figure (d), the STTs exert almost totally on B atoms close to MgO/Mg$_{4}$BO$_{3}$ interface, which could be attributed to the large magnetic moments and the evanescent states on ferromagnetic impurity B atoms. 
  }\label{layer}
\end{figure}

\begin{figure*}[tbh]
  \includegraphics[width=18cm]{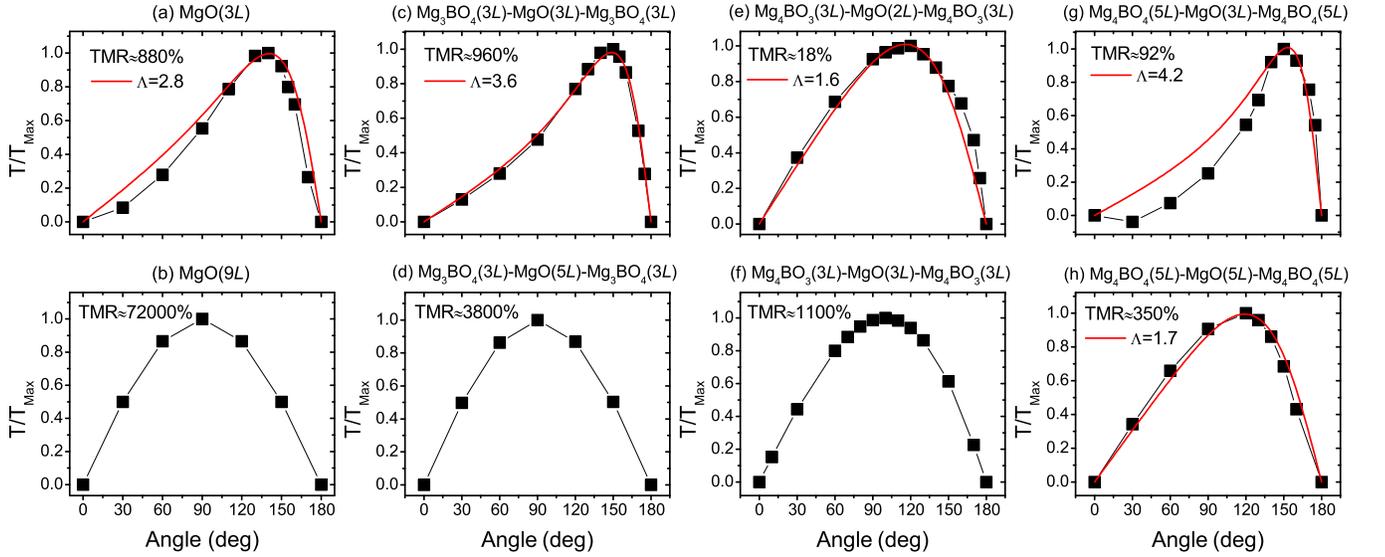}\\
  \caption{(color online). The angular dependence of total in-plane STTs in CoFe/MgO and CoFe/Mg-B-O based MTJs with different spacers labeled in the titles of the figures. The red solid lines are the fitted Slonczewski's formulas with asymmetry parameter $\Lambda$. As shown in figure (c) and (g), a strong asymmetry of angular dependence of STT is obtained in CoFe/Mg$_{3}$BO$_{4}$(3\emph{L})/MgO(3\emph{L})/Mg$_{3}$BO$_{4}$(3\emph{L})/CoFe and CoFe/Mg$_{4}$BO$_{4}$(5\emph{L})/MgO(3\emph{L})/Mg$_{4}$BO$_{4}$(5\emph{L})/CoFe junctions respectively. Such asymmetry of STTs could be attributed to the existence of interfacial resonance states induced by B atoms in Mg-B-O spacer. 
  }\label{stt}
\end{figure*}

\section{Results and Discussion}

Firstly, the density of state (DOS) for three types of bulk B-doped (25\%) MgO crystals are calculated. Unlike the initial undoped MgO, all of the three B-doped MgO crystals exhibit metallic state, as shown in Fig.~\ref{dos}. The present calculations also show that the bulk Mg$_{3}$BO$_{4}$ is paramagnetic. Meanwhile, the B atoms in Mg$_{4}$BO$_{3}$ have large nonzero magnetic moments ($\sim2.0\mu_{B}$), consisting with the previous first-principles calculations.~\cite{Liu-jpc10, Bannikov-tpl07} The magnetic moment of B in Mg$_{4}$BO$_{4}$ is relative small ($\sim0.15\mu_{B}$) in comparison with that of ordered CoFe lead ($\mu_{\text{Co}}\sim$1.7$\mu_{B}$, $\mu_{\text{Fe}}\sim$2.7$\mu_{B}$). Moreover, the shaded parts in Fig.~\ref{dos} display the projected DOS on impurity B atoms. Clearly, a significant coupling between the impurity B and MgO atoms at Fermi level appears in Mg$_{3}$BO$_{4}$ and Mg$_{4}$BO$_{4}$, whereas such coupling is much smaller in Mg$_{4}$BO$_{3}$. As a result, this difference of projected DOS should be responsible for the different STT characters among the three types of CoFe/Mg-B-O/CoFe MTJs.

Fig.~\ref{layer} presents the calculated layer resolved in-plane STTs of MTJs at unit bias with [Fig.~\ref{layer}(c)-(e)] and without [Fig.~\ref{layer}(a) and (b)] B diffusion for the relative angle of 90 degree between two leads. Here the calculations show that the torques emerge mainly on Co or Fe atoms near the CoFe/Mg-B-O interface, similar to the first-principles calculations in Fe/MgO system.~\cite{jia-prl11, stiles-prl08} However, in CoFe/Mg$_{4}$BO$_{3}$ based MTJ the STTs on B atoms which is close to MgO/Mg$_{4}$BO$_{3}$ interface are much larger than that on Co or Fe atoms [see Fig.~\ref{layer}(d)]. This behavior is partially due to the large magnetic moments on those B atoms ($\sim2.0\mu_{B}$) and the metallic state of B-doped MgO, which make the B-doped MgO layer become a interfacial part of ferromagnetic electrodes of MTJ. So the spin current will be absorbed in this region leading to angular momentum transfer torques. Furthermore, the evanescent state in ferromagnetic impurity B could also contribute to the STT, which is responsible for the rapid vanishing of the torque away from B atoms. In addition, when the bias is applied on CoFe/Mg$_{4}$BO$_{3}$ based MTJ, although the torques exert almost totally on B atoms, the magnetization of CoFe lead could still be switched by current owing to the possible exchange interaction between B and CoFe lead. Therefore, the total in-plane STT calculated in this work means the sum of the in-plane STTs on all atoms (including magnetic B atoms) at the right half of MTJ.

Fig.~\ref{stt} gives the angular resolved total in-plane STTs for CoFe/MgO and CoFe/Mg-B-O based MTJs with different spacers. The highly asymmetric anglular dependence of STTs in CoFe/Mg$_{3}$BO$_{4}$ and CoFe/Mg$_{4}$BO$_{4}$ based MTJs are clearly shown in Fig.~\ref{stt}(c) and (g). Their corresponding atomic structures are presented in Fig.~\ref{structure}(a) and (c) respectively. Furthermore, the respective fitted Slonczewski parameters are $\Lambda(\text{CoFe/Mg$_{3}$BO$_{4}$})\approx3.6$ and $\Lambda(\text{CoFe/Mg$_{4}$BO$_{4}$})\approx4.2$, which suggests that the doping of B into MgO spacer of CoFe/MgO MTJs may be benefit to improve output power of high-frequency generation. Meanwhile, as shown in Fig.~\ref{stt}(f), the angular dependence of STT is a symmetric sine function for the CoFe/Mg$_{4}$BO$_{3}$ based MTJ whose structure is shown in Fig.~\ref{structure}(b). This STT curve is similar with the case of undoped CoFe/MgO/CoFe MTJ with 9 layers (9\emph{L}) MgO [see Fig.~\ref{stt}(b)], which indicates that the B atoms substituting O atoms in MgO spacer is not an effective way to get skewed STT. In addition, the previous study on formation energy~\cite{Chandra-prb14} observe that the B atom is favored to occupy the position of Mg atom in MgO, namely, the MTJ based on CoFe/Mg$_{3}$BO$_{4}$ should be the most likely formed. Therefore, we demonstrate that the asymmetry parameter $\Lambda$, which dominates the output power of high-frequency generation, could be increased after B moving into MgO spacer in the fabrication procedure of CoFe/Mg-B-O/CoFe MTJ.

To understand the angular asymmetry of STT, we also calculate the STT of CoFe/MgO(3\emph{L})/CoFe MTJ with 3 layers MgO spacer, as shown in Fig.~\ref{stt}(a). Obviously, its STT curve is highly asymmetric, similar to the results of Fe/MgO(3\emph{L})/Fe MTJ.~\cite{jia-prl11} So far, it is believed that the asymmetric character of STT in Fe/MgO MTJs with thin barrier are originated from the multiple scattering caused by the interfacial resonance states. Considering the B-doped MgO layers in spacer behave as conductive metal, the effective tunneling barrier thickness in CoFe/Mg-B-O MTJs are therefore decreased in comparison to the case of undoped CoFe/MgO MTJs. As a result, the interfacial states at Mg$_{3}$BO$_{4}$(Mg$_{4}$BO$_{3}$, Mg$_{4}$BO$_{4}$)/MgO interface can resonate once the effective barrier thickness is small enough.

\begin{figure}
  \includegraphics[width=8.6cm]{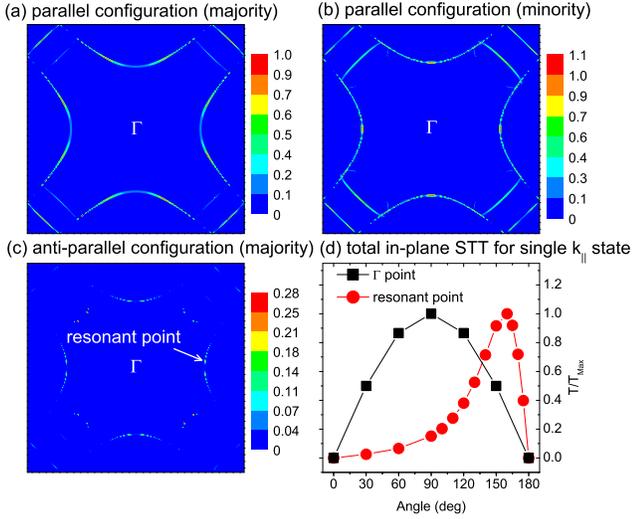}\\
  \caption{(color online). $\textbf{k}_\|$ dependence of transmittance for (a) majority, (b) minority of parallel configuration, and (c) majority of anti-parallel configuration for CoFe/Mg$_{3}$BO$_{4}$(3\emph{L})/MgO(3\emph{L})/Mg$_{3}$BO$_{4}$(3\emph{L})/CoFe junction. Figure (d) presents the angular dependence of STTs for $\Gamma$ and one of the resonant point highlighted by an arrow in figure (c). The angular dependence of STT is very skewed for this resonant $\textbf{k}_\|$ state, as shown in figure (d). }\label{irs}
\end{figure}

As we well know, the interfacial resonance state can be exponentially suppressed when the barrier thickness is increasing. In the MTJs with the structures shown in Fig.~\ref{structure}(a)-(c), the barrier thickness is determined by the number of layers of middle undoped MgO. Therefore we further calculated the STTs of CoFe/Mg$_{3}$BO$_{4}$(3\emph{L})/MgO(5\emph{L})/Mg$_{3}$BO$_{4}$(3\emph{L})/CoFe junction which has 5 layers of undoped MgO in the middle of spacer. As shown in Fig.~\ref{stt}(d), the obtained angular dependence of STT recovers to a symmetric sine function, consisting with the fact of exponentially suppressed behavior. Similarly, when we insert 5 layers of undoped MgO in the middle of CoFe/Mg$_{4}$BO$_{4}$ based MTJ, the angular dependence of STT for such junction of CoFe/Mg$_{4}$BO$_{4}$(5\emph{L})/MgO(5\emph{L})/Mg$_{4}$BO$_{4}$(5\emph{L})/CoFe also becomes more symmetric [see Fig.~\ref{stt}(h)] despite the asymmetry parameter is still larger than 1 ($\Lambda\approx1.7$), which is due to the effective barrier thickness is failed to decouple the interfacial states here.

\begin{figure}
  \includegraphics[width=8.6cm]{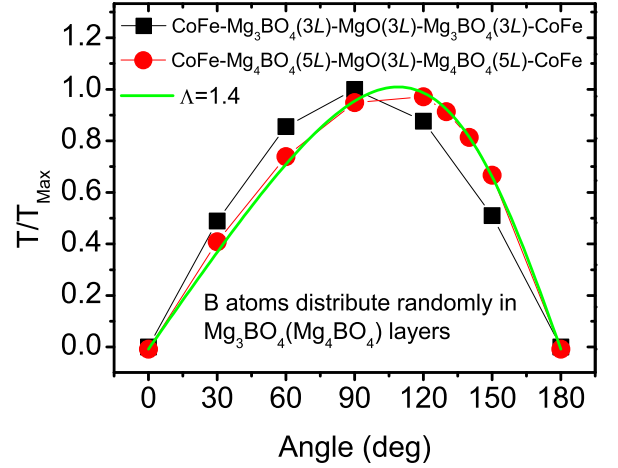}\\
  \caption{(color online). The angular dependence of total in-plane STTs for disordered CoFe/Mg$_{3}$BO$_{4}$ and CoFe/Mg$_{4}$BO$_{4}$ based MTJs with B atoms distributing randomly in Mg$_{3}$BO$_{4}$ and Mg$_{4}$BO$_{4}$ layers respectively. The green solid lines is the fitted Slonczewski's formula with asymmetry parameter $\Lambda=1.4$. The results indicate that the disorder of B atoms distribution could effectively restore the angular dependence of STT to symmetric sine function.   
  }\label{disorder}
\end{figure}

\begin{figure*}[tbh]
\includegraphics[width=16cm]{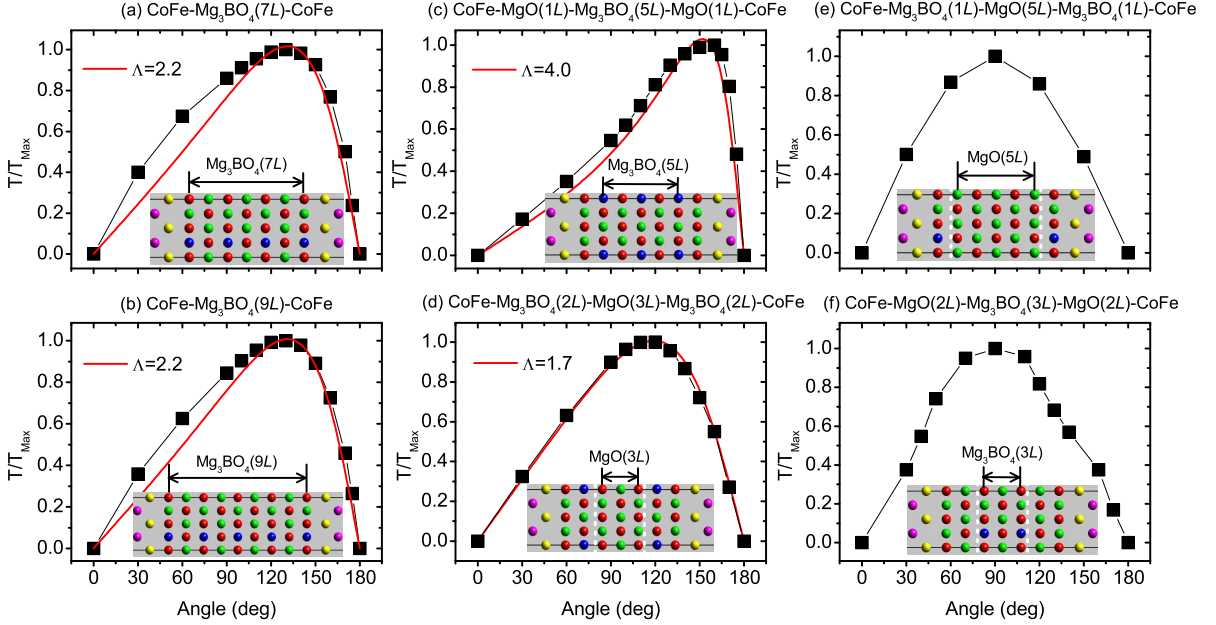}
\caption{(color online). The angular dependence of total in-plane STTs for several configurations of CoFe/Mg$_{3}$BO$_{4}$ based MTJs, whose atomic structures are plotted in the insets. The red solid lines are the fitted Slonczewski's formulas with asymmetry parameter $\Lambda$. As shown in figure (a)-(c), the angular dependent of STTs are still highly asymmetric for the MTJs with B atoms diffusion into the entire region of spacer. Due to the nature of metallic state at Fermi level in Mg$_{3}$BO$_{4}$ spacer, the high spin polarization of CoFe lead is responsible for the skewness of STT, similar with the case of spin valve with half-metallic ferromagnetic electrodes. 
  }\label{mg3bo4}
\end{figure*}

To reveal interfacial resonances in more detail, in Fig.~\ref{irs}(a)-(c) we present the $\textbf{k}_\|$ dependence of transmittance for CoFe/Mg$_{3}$BO$_{4}$(3\emph{L})/MgO(3\emph{L})/Mg$_{3}$BO$_{4}$(3\emph{L})/CoFe with parallel (P) and anti-parallel (AP) magnetic configuration. It can be seen clearly that the hot-spots of conductance owing to interfacial resonance states appear within 2D BZ for both P and AP configuration. Moreover, the results show that the $\textbf{k}_\|$ states with resonant tunneling for majority and minority of P configuration are nearly in the same positions within 2D BZ. Meanwhile, the resonant points of AP configuration are also contributed almost by these $\textbf{k}_\|$ states. Here we find that the angular dependence of STT is quite different between the resonant state and non-resonant state. Take Fig.~\ref{irs}(d) as an example, it gives the angular dependence of STT for one of this resonant tunneling $\textbf{k}_\|$ states which is highlighted by an arrow in Fig.~\ref{irs}(c). Such angular dependence of STT is very skewed for this resonant tunneling state, as shown in Fig.~\ref{irs}(d). On the other hand, to compare the case of resonant state, we calculate the STT of $\Gamma$ point also plotted in Fig.~\ref{irs}(d) and found that the STT curve is symmetric sine function. Because the interface transparency dominates the magnitude of STT,~\cite{tang-jap13} the resonant point within 2D BZ contributes mainly to the torque of MTJ due to its large conductance. As a result, the very skew STT curve could be obtained when the significant interfacial resonance emerges in MTJ.

However, unlike the case of CoFe/Mg$_{3}$BO$_{4}$ based MTJ, the angular dependence of STT of CoFe/Mg$_{4}$BO$_{3}$ based MTJ whose structure is shown in Fig.~\ref{structure}(b) is still symmetric, as shown in Fig.~\ref{stt}(f). Here we argue that the interfacial resonance is greatly weaker in CoFe/Mg$_{4}$BO$_{3}$ based MTJ than that in CoFe/Mg$_{3}$BO$_{4}$ based MTJ, although both MTJs all have 3 layers undoped MgO in the middle of spacer. For the MTJs with the structures plotted in Fig.~\ref{structure}(a) and (b), the interfacial resonance states mainly come from the states on B atoms at the MgO/Mg$_{3}$BO$_{4}$ or MgO/Mg$_{4}$BO$_{3}$ interface. Therefore, the strength of corresponding interfacial resonance is proportional generally to the coupling between such interfacial B atoms and the undoped MgO layers in the middle of spacer. Furthermore, from Fig.~\ref{dos}(a)-(c) one can infer that at Fermi level the orbital admixture between the B and MgO is relative weaker for Mg$_{4}$BO$_{3}$ crystal. Therefore, the coupling between B atoms and the middle undoped MgO atoms in CoFe/Mg$_{4}$BO$_{3}$ based MTJ is certainly weaker. Accordingly, the corresponding interfacial resonance will be much more difficult to be built compared to the case of CoFe/Mg$_{3}$BO$_{4}$ based MTJ under the same thickness of middle undoped MgO layers.

With the thickness of middle undoped MgO layers decreasing, the interfacial resonance in CoFe/Mg$_{4}$BO$_{3}$ based MTJ can be restored. To confirm the existence of interfacial resonance state in CoFe/Mg$_{4}$BO$_{3}$ based MTJ, we will remove one layer of undoped MgO in the middle of the MTJ whose structure is plotted in Fig.~\ref{structure}(b) and calculate the STT of CoFe/Mg$_{4}$BO$_{3}$(3\emph{L})/MgO(2\emph{L})/Mg$_{4}$BO$_{3}$(3\emph{L})/CoFe junction, which has only 2 layers undoped MgO in the middle of spacer. Note that this kind of MTJ is no longer a symmetric junction, i.e., the interface between right lead and spacer is Co-Mg instead of Co-O. As expected, the STT curve indeed starts to become skew at the case of only 2 layers middle undoped MgO, as shown in Fig.~\ref{stt}(e). Hence, the present results suggest that the resonance behavior of interfacial states also occur in the CoFe/Mg$_{4}$BO$_{3}$ based MTJ, although it is relatively weaker than that of CoFe/Mg$_{3}$BO$_{4}$ based MTJ.

Moreover, the calculated TMR ratio also reflects the existence of interfacial resonance states in CoFe/Mg-B-O MTJs. As presented in Fig.~\ref{stt}, the B atoms diffusion into the MgO spacer will decrease TMR ratio compared to the case of undoped CoFe/MgO MTJs. In addition, with the thickness of middle undoped MgO layer increasing, the TMR ratio will recover to a larger value. Such behaviors of TMR ratio can be attributed to the reduction of effective tunneling barrier thickness due to the metallic proprieties of B-doped MgO layer. Therefore, when the barrier thickness is cut down to very thin ($\sim$ 3 layers), the emergence of interfacial resonance will reduce the TMR ratio, similar to the case of Fe/MgO MTJ. ~\cite{Belashchenko-prb05}

Next, we investigate how the random distribution of B atoms influences the skewness of STTs in CoFe/Mg-B-O MTJs. Here we consider the CoFe/Mg$_{3}$BO$_{4}$ and CoFe/Mg$_{4}$BO$_{4}$ based MTJs with the structures shown in Fig.~\ref{structure}(a) and (c) respectively. In order to simulate the random distribution, the 3$\times$3 lateral supercell without distortion is constructed. The B atoms randomly substitute for Mg atoms in Mg$_{3}$BO$_{4}$ layers of CoFe/Mg$_{3}$BO$_{4}$ based MTJ. Likewise, the B atoms also randomly occupy the interstitial positions in Mg$_{4}$BO$_{4}$ layers of CoFe/Mg$_{4}$BO$_{4}$ based MTJ. In the present transport calculations the frozen potential model is used, namely the potential of disordered 3$\times$3 lateral supercell is constructed by randomly distributing the self-consistent atomic sphere potential of B and Mg atoms which are obtained from the case of ordered supercell. In Fig.~\ref{disorder}, the results show that the angular dependence STT of CoFe/Mg$_{3}$BO$_{4}$ based MTJ recovers to symmetric ($\Lambda=1$) when B atoms distribute randomly. Similarly, the degree of symmetry for STT curve in disordered CoFe/Mg$_{4}$BO$_{4}$ based MTJ is larger than that in ordered MTJ, although it is still a skew curve with $\Lambda\approx1.4$. According to the above results, the disorder of B atoms distribution will destroy the interfacial resonance states, which is the origin of asymmetric angular dependence of STT. Obviously, to improve the output power of STO based on CoFe/Mg-B-O MTJs, the B atoms need to be crystallized into a ordered lattice in Mg-B-O spacer during the preparation of junctions.

To date, it is known that the B atoms have diffused into the MgO spacer but where and how these B atoms exactly distribute are still unclear yet. Thus we further demonstrate the angular dependence STTs for several different configurations of CoFe/Mg$_{3}$BO$_{4}$ based MTJs, in which the thickness of spacer is about 1.4 nm.

Firstly, the MTJs with B atoms diffusion into the entire region of spacer are considered. As shown in Fig.~\ref{mg3bo4}(a), for CoFe/Mg$_{3}$BO$_{4}$(7\emph{L})/CoFe we found that the STT is significantly asymmetric with $\Lambda\approx2.2$. When the thickness of spacer is increasing, as shown in Fig.~\ref{mg3bo4}(b), the angular dependence STT of CoFe/Mg$_{3}$BO$_{4}$(9\emph{L})/CoFe is still asymmetric. Due to Mg$_{3}$BO$_{4}$ behaves as conductive metal, the whole system therefore could be viewed as the metallic multilayers spin valve. Thus, similar to the case of spin valve with half-metallic ferromagnetic lead,~\cite{tang-jap13} the high spin polarization of CoFe lead is responsible for the skewness of STT. In addition, this phenomenon is also found in CoFe/MgO(1\emph{L})/Mg$_{3}$BO$_{4}$(5\emph{L})/MgO(1\emph{L})/CoFe, in which the corresponding angular dependence of STT is very skew with $\Lambda\approx4.0$, as shown in Fig.~\ref{mg3bo4}(c).

Secondly, the CoFe/Mg$_{3}$BO$_{4}$ based MTJs with B atoms only diffusion into the edge of spacer near the left and right CoFe leads are considered. As expected, the calculations show that the angular dependence of STT is asymmetric ($\Lambda\approx1.7$) for the MTJ with 3 layers undoped middle MgO [see Fig.~\ref{mg3bo4}(d) for details]. Meanwhile, the STT curve is symmetric for the MTJ with 5 layers undoped middle MgO [see also Fig.~\ref{mg3bo4}(e) for details], which suggests that the interfacial states on Mg$_{3}$BO$_{4}$/MgO could't efficiently resonate at such large distance ($\sim$ 1 nm).

Finally, the MTJ with the B atoms distributing in the middle region of spacer are also investigated. As shown in Fig.~\ref{mg3bo4}(f), the maximum value of angular dependent STT occurs around 90 degree corresponding to Slonczewski's formula with $\Lambda=1$. This result shows that the STT behavior originated from interfacial resonance states is absent in the dopant form of B atoms locating at the middle region of spacer. As shown in the inset of Fig.~\ref{mg3bo4}(f), from the structure of this kind of MTJ one can infer that here the middle B-doped MgO layers could't couple the two CoFe/MgO interfacial states and the distance between two CoFe/MgO interface is too large to set up interfacial resonance efficiently. In other words, the B atoms residing in the region near CoFe lead is crucial to reduce the distance between two interfacial states and obtain the skew STT curve in such MTJ with 1.4 nm thick spacer. This conclusion could guild the experimental researchers to tune the B atoms occupancy accurately and thus increase the experimental output power of STO.

\section{Summary}

In summary, the angular dependence of spin-transfer torques in different configurations of CoFe/Mg-B-O/CoFe magnetic tunnel junctions are studied by a first-principles noncollinear wave-function-matching method. It is believed that the B atoms have diffused into MgO spacer after annealing process, and therefore three types of B-doped (25\%) MgO are considered here, (1) B substituting for Mg (Mg$_{3}$BO$_{4}$), (2) B substituting for O (Mg$_{4}$BO$_{3}$), and (c) B occupying an interstitial position (Mg$_{4}$BO$_{4}$) in MgO. A strong asymmetric angular dependence of STT with large asymmetry parameter $\Lambda$ could be obtained both in ballistic CoFe/Mg$_{3}$BO$_{4}$ and CoFe/Mg$_{4}$BO$_{4}$ based MTJs respectively. However, the skewness of STT can be suppressed by the disorder of B distribution in MgO spacer. Moreover, a relatively symmetric angular dependence of STT is observed in CoFe/Mg$_{4}$BO$_{3}$ based MTJ. These phenomenons could be attributed to the interfacial resonance states induced by B diffusion into MgO spacer. Our calculations demonstrate that the B atoms diffusion into MgO spacer in ballistic CoFe/Mg-B-O/CoFe MTJs is beneficial to produce high microwave output power in spin transfer oscillator.

\textbf{Acknowledgements} The authors are grateful to Prof. Ke Xia, Dr. Shuai Wang, and Dr. Yuan Xu for their preliminary works. We are also grateful to Dr. Anton Starikov for the TB-MTO code based upon sparse matrix techniques. This work is supported by the Natural Science Foundation of China (Grant No: 11304279, 11364007), the Natural Science Foundation of Zhejiang Province, China (Grant No: LY13A040004, LY16A040013), the China Postdoctoral Science Foundation (Grant No. 2012M520666), the Science and Technology Foundation from Ministry of Education of Liaoning Province (Grant No: L2015333), and the Science and Technology Foundation from Guizhou Province (Grant No: J [2013]2242).

\end{document}